\documentstyle[12pt]{article}
\begin{document}
\def\thebibliography#1{\section*{REFERENCES\markboth
 {REFERENCES}{REFERENCES}}\list
 {[\arabic{enumi}]}{\settowidth\labelwidth{[#1]}\leftmargin\labelwidth
 \advance\leftmargin\labelsep
 \usecounter{enumi}}
 \def\newblock{\hskip .11em plus .33em minus -.07em}
 \sloppy
 \sfcode`\.=1000\relax}
\let\endthebibliography=\endlist

\hoffset = -1truecm \voffset = -2truecm

\title{\large\bf
T-Dependent Dyson- Schwinger Equation In IR Regime Of QCD: The
Critical Point }
\author{
{\normalsize \bf A.N. Mitra$^\star$ \thanks{e.mail:
ganmitra@nde.vsnl.net.in} \quad and W-Y. P. Hwang \thanks{e.mail:
wyhwang@phys.ntu.edu.tw}
}\\
\normalsize  Center for Academic Excellence on
Cosmology \& Particle Astrophysics,\\
 National Taiwan University,\\
Taipei 106, Taiwan, R.O.C. \\
$^\star$Permanent address: 244 Tagore Park, Delhi-110009, India }
\date{}
\maketitle
\begin{abstract}
The quark mass function $\Sigma(p)$ in QCD is revisited, using a
gluon propagator in the form $1/(k^2 + m_g^2)$ plus $2\mu^2/ (k^2
+ m_g^2)^2$, where the second (IR) term gives linear confinement
for $m_g = 0$ in the instantaneous limit, $\mu$ being another
scale. To find $\Sigma(p)$ we propose a new (differential) form of
the Dyson-Schwinger Equation (DSE) for $\Sigma(p)$, based on an
infinitesimal $subtractive$ Renormalization  via a differential
operator which $lowers$ the degree of divergence in integration on
the RHS, by $TWO$ units. This warrants $\Sigma(p-k)\approx
\Sigma(p)$ in the integrand since its $k$-dependence is no longer
sensitive to the principal term $(p-k)^2$ in the quark propagator.
The simplified DSE (which incorporates WT identity in the Landau
gauge) is satisfied for large $p^2$ by $\Sigma(p)$ = $\Sigma(0)/(1
+ \beta p^2)$, except for Log factors. The limit $p^2 =0$
determines $\Sigma_0$.A third limit $p^2 = -m_0^2$  defines the
$dynamical$ mass $m_0$ via $\Sigma(im_0) = + m_0$. After two
checks ($f_\pi = 93\pm 1 MeV $ and $ <q{\bar q}>$= $(280 \pm 5
MeV)^3$), for $1.5<\beta<2$ with $\Sigma_0=300 MeV$, the T-
dependent DSE is used in the real time formalism to determine the
"critical" index $\gamma= 1/3$ analytically, with the IR term
partly  serving for the $H$ field. We find $T_c = 180 \pm 20 MeV$
and check the vanishing of $f_\pi$ and $<q{\bar q}>$ at $T_c$.
\end{abstract}

\section{Introduction}

QCD, as the queen of strong interaction theory, lies at the root
of a whole complex of strong interaction phenomena, ranging from
particle physics to cosmology.Its principal tool is the quark mass
function, termed $\Sigma(p)$ in the following, as a central
ingredient for the evaluation of  a string of QCD parameters whose
primary examples are the pion decay constant and the quark
condensate. The thermal behaviour of the latter in turn  has
acquired considerable cosmological relevance in recent years in
the context of global experimentation on heavy ion collisions as a
means of accessing the quark-gluon plasma (QGP) phase [1- 4]. It
is therefore essential to have on hand a reliable $\Sigma(p)$
function in a non-perturbative form as a first step towards the
evaluation of these basic QCD parameters. In this respect, QCD sum
rule (SR), attuned to finite temperatures [2] have been  a leading
candidate for such studies for a long time, using the $FESR$
duality principle [5], as well as a variational approach via the
minimum of effective action up to two loops (Barducci et al, [1a])
to determine the mass function. An alternative approach has been
the method of chiral perturbation theory [6] with the $pion$ as
the basic unit in preference to quarks. Now a standard approach to
QCD is via the RG equation  for the $\beta$ function in the lowest
order of $g$ which yields
 $$\alpha_s (Q^2) = 2\pi /[9 \ln {(Q /\Lambda_Q)}] $$
 with 3 flavours, $\Lambda_Q$ being the QCD scale parameter[7].
 Unfortunately the higher order terms in $g$ are not particularly amenable
 to the simulation of non-perturbative effects. On the other hand,
 the Dyson-Schwinger Equation (DSE) which may be regarded as the differential
 form of the minimum principle of effective action [8], offers a
 more promising tool  which has often  been  used with the standard
 o.g.e. in the rainbow approximation [9], but can be improved to
 incorporate gauge invariance so as to satisfy the W-T identity in
 the "dynamical perturbation theory" (which ignores cris cross gluon
 lines in the skeleton diagrams) with little extra effort, as first
 shown by Pagels-Stokar [10]. In this paper, we shall use the same
 approach, but  explicitly add an extra, non-perturbative, term to
 the one-gluon-exchange (oge) propagator
 for a quicker simulation of the infrared (IR) regime, so that both
 together act as the `kernel' of the Dyson Schwinger Equation (DSE)
 [11]. Thus the total gluon propagator is given by

\setcounter{equation}{0}
\renewcommand{\theequation}{1.\arabic{equation}}

\begin{equation}\label{1.1}
G(k) = \frac{1}{k^2 + m_g^2} + \frac{2\mu^2}{k^2 + m_g^2)^2}
\end{equation}
where $\mu$ is a scale parameter corresponding to the (hadronic)
$GeV$ regime, (whose value will be left undetermined till later),
and $m_g$ is a (small) gluon mass a non-zero value for which can
be motivated from several angles, a notable one being the
`Schwinger mechanism' [12] as explained in the Jackiw-Johnson
paper [13]. A second motivation was highlighted by Cornwall et al
[14], in the context of their approach to a more compact
realization of gauge invariance via the so-called `pinch
mechanism' [14]. Yet a third motivation which is especially
relevant in the present context of a temperature dependent DSE,
comes from according it a `Debye mass' status, running with the
temperature [15]. A non-perturbative gluon propagator (with
harmonic confinement)  was  employed in [16-17] as a kernel of a
BSE for the $gg$ wave function for the calculation of glueball
spectra , on similar  lines to $q{\bar q}$ spectroscopy [18]
Alternative BSE treatments for glueballs also exist in the
literature [19]. In this paper, the IR part of Eq.(1.1) has a dual
role: 1) to serve as a more efficient simulation of the
non-perturbative effects on the mass function $\Sigma(p)$ ; and 2)
a partial simulation of the external magnetic field effect, as an
alternative to small non-zero masses of `current' quarks [20,15].
With a non-perturbative solution of the DSE, we are primarily
concerned with chiral symmetry restoration at a critical
temperature $T_c$. To that end we shall be interested in the $T$
behaviour near the critical Point $T_c$, rather than as an
expansion in powers of $T^2$ near $T=0$ [6]. Note however that
linear confinement ($\sim r$) corresponds to $m_g = 0$, via the
second term in (1.1), in the (3D) instantaneous limit $t=0$, so
that deconfinement competes with chiral symmetry restoration with
a propagator like (1.1). We shall not pursue this aspect further,
although we note that deconfinement has been claimed to occur at a
lower temperature [21] than chiral symmetry restoration.

\subsection{Object and Scope of The Paper}

The central object of this paper is a determination of the mass
function $\Sigma(p)$ non-perturbatively in the intermediate
momentum regime with the help of  the gluon propagator (1.1) that
covers the IR regime. This is sought to be achieved via a (new)
differential formulation of the DSE based on a subtractive form of
renormalization that is particularly convenient for a DSE type
equation. A second object is to apply the $\Sigma(p)$ so
determined, to two basic quantities, $<q{\bar q}>$ and $f_\pi$,and
express them in an $analytic$ form, so that their $T$- dependent
generalizations may be achieved analytically too. A third object
is to generalize the DSE to a $T$-dependent form, so as to obtain
an equation for the $T$-dependent mass function $m_t$, with a
focus on its critical index $\gamma$ associated with the critical
temperature $T_c$, so as to gauge the role of the IR term
vis-a-vis small current masses to simulate the $H$-field effect
[20, 15]. Further, while in the conventional methods [20, 15], the
various thermodynamic quantities are derived from a central
quantity like the free energy [15], or equivalently the effective
potential [20], and taking appropriate derivatives, the plan
adopted here is to focus on the DSE itself as the principal form
of dynamics, with  $m_t$ as a natural order parameter. Due to the
unconventional nature of this approach, this part of the exercise
is still preliminary, with only one critical index identifiable
with $m_t$ determined so far, while other indices [20] are left
for later studies, within the DSE framework.
\par
 In {\bf Sect.2}, we formulate the DSE for $\Sigma(p)$
in an (infinitesimal) form of (subtractive) renormalization which
yields a non-linear second order differential equation for this
quantity. The $dynamical$ mass $m_0$  is defined as  the pole of
$S_F(p)$ at $ i\gamma. p = - m_0$, and  hence corresponds  to the
solution of the equation for $ \Sigma(i m_0) = m_0$. Although in
principle a mass renormalization factor $Z_m$ comes according to
rules [7], the condition $ \Sigma(i m_0) = m_0$  ensures that this
factor is effectively unity, provided the dynamical mass is
employed for the propagator at its pole. As for the quantity
$\Sigma(0)$, we shall designate it as the $constituent$ mass. For
the solution of the resulting DSE, three crucial check-points are
$p^2 = \infty$; $p^2 = 0$; $p^2 =- m_0^2$ which control the
structure of $\Sigma (p)$. The simplest ansatz consistent with a
$p^{-2}$ - like behaviour in the $p^2 = \infty $ limit, as
demanded by QCD, is $ \Sigma_0 / (1 + \beta p^2) $ [1], the only
precaution needed for a consistent solution being a constant
$\alpha_s$ with its argument $fixed$ in advance at a certain
specified value. This form has good analytical properties for
large space-like momenta, but it implies that the dynamical mass
$\Sigma(im_0)$ exceeds the constituent mass $\Sigma(0)$.
\par
For a basic test of this structure, we choose in {\bf Sect.3},two
key items i) $q{\bar q}$ and ii) $f_\pi^2$  whose derivations are
sketched in Appendices A and B respectively in an $analytical$
form. The results agree with experiment to within $\sim 5 \%$, for
$\Sigma_0 =300 MeV$, $m_g \approx \Lambda_Q = 150 MeV$, and the
hadronic scale parameter $\beta$ in the range $1.5 < \beta < 2.0$.
\par
{\bf Sect.4}  outlines the formulation of the temperature
dependent DSE (T-DSE for short) within the real time formalism
[22], instead of the imaginary time formalism a la Matsubara [23].
 The order parameter in this regard may be
chosen in one or more ways, a convenient choice being $\Sigma_0$
which now "runs" with the temperature and is renamed as $m_t$.
Other analogous quantities which are expected to "run" with the
temperature are the gluon mass renamed as $m_{gt}$,  and perhaps
also the IR parameter $\mu$ whose connection with $m_t$ and
$m_{gt}$ is brought out in Sect.(4).  It is found that the both
the constituent and gluon masses have the $same$ "critical index"
$\gamma = 1 / 3$ (in accordance with the concept of `universality'
of critical indices), while the critical temperature works out at
$  T_c  \approx 180 \pm 20 MeV $. Sect.5 concludes with a
discussion including a comparison with contemporary approaches.

\section{Dyson-Schwinger Eq In Differential Form }

\setcounter{equation}{0}
\renewcommand{\theequation}{2.\arabic{equation}}

We start by writing the DSE in the Landau gauge which ensures that
the $A$ parameter does not suffer renormalization [24]. This is an
additional precaution over and above the Pagels-Stokar DPT
approach [10] to satisfy the WT identity. The starting DSE in the
Landau gauge for the function $\Sigma(p)$, after tracing out the
Dirac matrices takes the form
\begin{eqnarray}\label{2.1}
\Sigma(p) &=& ig_s^2 \frac{F_1.F_2}{(2\pi)^4}\int d^4k
\frac{1}{[\Sigma^2(p-k) + (p-k)^2]}\times  \\  \nonumber
          & &  [\Sigma(p-k)\delta_{\mu\nu}+ (\Sigma(p)-\Sigma
(p-k)) \frac{(p-k)_\mu (2p-k)_\nu}{k^2- 2p.k}]  \\  \nonumber
          & &  (\delta_{\mu\nu}- k_\mu k_\nu / k^2)
[\frac{1}{k^2 + m_g^2} + \frac{2\mu^2}{(k^2 + m_g^2)^2}]
\end{eqnarray}
The first term on the RHS corresponds to the rainbow approximation
[9], while the second term gives the simplest realization of a
gauge invariant structure  by satisfying the WT identity a la
Pagels-Stokar [10]. An analogous but slightly more involved ansatz
due to Ball-Chiu [25] also can be seen to conform  to the Landau
form [10], through a visual inspection of both. To see this more
explicitly, we list both forms for the relevant vertex functions,
first [25] (as given in [24]) followed by [10]:
\begin{eqnarray}\label{2.2}
\Gamma_\nu(p', p) &=& -i\gamma_\nu [A+A']/2 +
\frac{A'-A}{2(p^2-p'^2)}[-i\gamma.(p+p')(p_\nu+p_\nu')] +
\frac{B-B'}{p^2-p'^2}(p_\nu+p_\nu')  \nonumber \\
\Gamma_\nu(p', p) &=& -i\gamma_\nu
+\frac{\Sigma(p)-\Sigma(p')}{p^2-p'^2}(p_\nu+p_\nu')
\end{eqnarray}
where the momentum dependence ($p, p'$) of the Ball-Chiu functions
$A,B$ is indicated by the unprimed and primed notations
respectively and the mass function $\Sigma(p) = B/A$, while the
Landau gauge corresponds to $A=1$. The Ball-Chiu form [25] is seen
to be  compatible with Pagels-Stokar [10] (which is already in the
Landau gauge $A=1$). So, ithout further ado, we shall use only
[10] for simplicity.
\par
We now adopt a $subtractive$ form of Renormalization by writing a
similar equation for, say, $p'$, and $subtracting$ one from the
other. If $p'$ is infinitely close to $p$, this results in a
differential form. Thus we subject both sides of the eq.(2.1) to
the differential operator $ p.\partial $, $not$ the scalar form
$p^2\partial_{p^2}$, since the former  is more naturally attuned
to handling  $two$ vectors $p,k$ that occur on the RHS.  The main
advantage of this crucial step is to reduce the degree of
divergence of the integral w.r.t. $k$ by $two$ units, which in
turn allows further simplifications on $\Sigma (p-k)$ on the RHS,
since it falls off rapidly with $k^2$. In particular, we are
allowed the  following simplification as a result of this crucial
step of reducing divergence via differentiation:

$$ \frac{\Sigma(p) - \Sigma(p-k)}{k^2 - 2p.k} \approx -  \partial_{p^2}\Sigma(p) $$
A second simplification  arises from a contraction  of the factors
$(p-k)_\mu(2p-k)_\nu$ and $(\delta_{mu\nu} - k_\mu k_\nu / k^2)$
which is almost $independent$ of $k_\mu$, and gives on angular
integration [26] :
$$ 2[ p^2 - (p.k)^2 / k^2]  \approx  2 (1 - n^{-1})  p^2 ;  \quad  n =4  $$
Further, against the background of the differential operator
$p.\partial_p$ on both sides of (2.1), we can replace the mass
function $\Sigma^2(p-k)$ inside the fermion propagator on the RHS
due to an improved $k$-convergence, by simply replacing
$\Sigma^2(p-k)$ with $\Sigma^2(p)$, since this quantity already
falls off with momentum (see also [10]). The resulting eq.(2.1)
now takes the form
\begin{eqnarray}\label{2.3}
2\Sigma'(p) &=& \frac{4 g_s^2}{i(2\pi)^4} \int d^4k
[\frac{2\Sigma'(p) - \Sigma''(p)}{D(p-k)}
                    - \frac{\Sigma(p) - \Sigma'(p)/2}{D^2(p-k)}  \\  \nonumber
            & & (4\Sigma(p)\Sigma'(p) + 2p^2-2p.k)]
                [\frac{1}{k^2 + m_g^2} + \frac{2\mu^2}{(k^2 + m_g^2)^2}]
\end{eqnarray}
where we have taken $F_1.F_2 = -4/3$,  and defined derivatives and
propagators as
\begin{eqnarray}\label{2.4}
\Sigma'(p) &=& (1/2) p.\partial_p \Sigma(p) = p^2
\partial_{p^2}\Sigma(p); \nonumber  \\
 D(p-k)    &=& \Sigma^2(p) + (p-k)^2
\end{eqnarray}
Note that decoupling of $\Sigma(p)$ from $k_\mu$ now facilitates
the $k$-integration, thus converting the DSE into a differential
equation, while the form of $\Sigma(p)$ is as yet  undetermined.
(This structure is different from a more conventional one for a
differential form of the DSE, by making the $D(p-k)$ separable in
terms of $p_>$ and $p_<$, etc [7, 14]).
\par
The next task is to integrate w.r.t. $d^4k$ which for the o.g.e.
term is still logarithmically divergent and hence requires
`dimensional regularization ' (DR) a la t'Hooft-Veltmann [27],
while the IR term gives a convergent integral. We hasten to add
that this divergence (despite the Landau gauge) may well be an
artefact of the approximation $\Sigma(p-k) \approx \Sigma(p)$ in
the numerator on the RHS, but since the divergence is only
logarithmic, it is not sensitive to DR [27], and in any case it is
a small price to pay for the huge advantage accruing from this
(new) differentiation method for renormalization. Another
approximation concerns the factor $g_s^2$ on the RHS of eqs(2.1)
and (2.2) which, strictly speaking, is a function of the momenta
$p, k$, but at this stage we must  "freeze" the value of
$\alpha_s$ at a $fixed$ value (to be specified below) so as to get
a self-consistent asymptotic solution in the $p^2=\infty$ limit.
[More general solutions with the differential form (2.2) and
variable $\alpha_s$ have not been attempted here].

\subsection{Dimensional Regularization for Integrals}

Denote  the two  integrals  of Eq. (2.2) containing the o.g.e.
term only  by $I$ and $II$ respectively, of which only $I$ is
divergent (see above), but $II$ is convergent by itself. .Thus
write for $I$ in the Euclidean notation for dimension $n$, using
the DR method [27, 11]
\begin{eqnarray}\label{2.5}
I &=& 4 g_s^2 (2\Sigma'(p)- \Sigma''(p))\zeta^\epsilon \int
\frac{d^n k}{(2\pi)^n D(p-k)(k^2 + m_g^2)}  \\  \nonumber
  &=& 4 g_s^2 (2\Sigma'(p)- \Sigma''(p))\zeta^\epsilon \int_0^1 du \int_0^\infty d k^2 k^{n-2}
\frac{\pi^{n/2}}{\Gamma(n/2) (2\pi)^n (\Lambda_u + k^2)^2}
\end{eqnarray}
 where we have introduced the  Feynman variable $0 \leq u \leq 1$,
 $\zeta$ is a UV dimensional constant,  $\epsilon = 4-n$,  and
$$ \Lambda_u = u \Sigma^2(p) + p^2 u(1-u) + m_g^2 (1-u)  $$
The integration over $k^2$  is now straightforward , while that
over $u$ is simplified by dropping the $m_g^2$ term since there is
no infrared divergence.The result of all these steps after
subtracting the UV divergence [27] is (with $g_s^2 = 4\pi
\alpha_s$):
\begin{equation}\label{2.6}
I= (\alpha_s/\pi)(2\Sigma'(p)- \Sigma''(p))  [ \ln 4\pi - \gamma
+1 + \ln (\zeta^2 / A_p)]
\end{equation}
where
\begin{equation}\label{2.7}
A_p \equiv  \Sigma^2(p) + p^2 /2
\end{equation}
The other integral $II$ which is UV convergent, does not need DR
[27] and gives
\begin{equation}\label{2.8}
II = -(\alpha_s/ \pi)(\Sigma(p)-
\Sigma'(p)/2)\frac{(4\Sigma(p)\Sigma'(p)+ p^2)}{A_p}
\end{equation}
Thus the resulting DSE  may be expressed compactly from (2.3) as
\begin{equation}\label{2.9}
2\Sigma'(p) = I +II + I' + II'
\end{equation}
where we have taken $I$ and $II$ from the o.g.e. contributions
(2.7) and (2.8) respectively, as well as  added two
similar(infrared)) terms $I'$ and $II'$ arising from the second
(IR) part of the gluon propagator (1.1). In the same normalization
as above the last two work out as
\begin{eqnarray}\label{2.10}
I'  &=& \frac{2\mu^2 \alpha_s}{ \pi A_p}(2\Sigma'(p)-
\Sigma''(p))[\ln (A_p / m_g^2) -1] \\  \nonumber II' &=&
-\frac{2\mu^2 \alpha_s}{ \pi A_p^2}(\Sigma(p)- \Sigma'(p)/2)( 4
\Sigma(p)\Sigma'(p) + p^2)[\ln (A_p / m_g^2) -2]
\end{eqnarray}
where we have made use of the smallness of $m_g$ is simplifying
some of the integrals over $u$.  Note that the last two terms are
at least of two lower orders in $p$ than their o.g.e.
counterparts, so that they will not contribute to the $p^2 =
\infty$ limit of the differential equation (2.9).

\subsection{ Large and Small $p^2$ Limits of  DSE for $T=0$}

To solve eq.(2.9), we try the ansatz [1, 20]:
\begin{equation}\label{2.11}
\Sigma(p) = \Sigma_0 / [1 + \beta p^2]
\end{equation}
whose asymptotic  form is compatible with perturbative QCD
expectations for massless quarks in the chiral limit [10]. And
take the $fixed$ value of $p^2$ in the argument of $\alpha_s$ at
$p^2=\zeta^2$ where $\zeta$ is the UV parameter corresponding to
the upper limit of $p^2$ allowed in the solution of the DSE.
[Other options exist, but not particularly convenient].

\subsubsection{Large $p^2$ limit}
Remembering the definition (2.4) for $\Sigma'$, etc., we have in
the large $p^2$ limit for the  function (2.11),
$$  \Sigma (p) \approx  - \Sigma'(p)  \approx  + \Sigma''(p)  $$
Now remembering the upper limit of $p^2$ being constrained by the
UV parameter  $\zeta^2$, substitution from (2.9) yields the result
\begin{equation}\label{2.12}
 -2 = \alpha_s / \pi [-3 (\ln(4\pi) -\gamma +1 + \ln{2}) - 3 ]; \quad
\pi / \alpha_s = \frac{9}{2} \ln(\zeta / \Lambda_Q)
\end{equation}
$\Lambda_Q = 150 MeV$ being the usual QCD scale parameter. Thus
eq.(2.12) determines the value of the maximum momentum $\zeta$
within this approach,  and shows that  our formalism  does not
permit  $p^2 $ to exceed $\zeta^2$.  Unfortunately eq.(2.12),
which corresponds to the check point $p^2 = \infty$,  restricts
$\zeta$ to a rather low value :
\begin{equation}\label{2.13}
\zeta / \Lambda_Q = 1.5490 ;  \quad  \zeta = 0.706 GeV  only
\end{equation}
where the $MS$ scheme (not  ${\bar MS}$ [28] ) has been employed.

\subsubsection{Small $p^2$ limit}

Next we consider the small $p^2$ limit of eq.(2.9)  where for the
(fixed) argument of $\alpha_s$ we continue (for mathematical
consistency) to maintain the same value of $\alpha_s$
corresponding to $p^2 =\zeta^2$,  leading after straightforward
simplifications to
\begin{equation}\label{2.14}
C_0 + \ln {x_1 / x_0} - 3  +  1/x_0 + [I'+II'] = 9\ln
(\zeta/\Lambda_Q); \quad C_0 \equiv \ln {4\pi} -\gamma = 1.9538;
\end{equation}
where the dimensionless quantities are defined as
\begin{equation}\label{2.15}
x_1 \equiv \zeta^2 {\beta};\quad x_0  \equiv \Sigma_0 ^2 {\beta}
\end{equation}
Note that eq.(2.14) has a big term on the RHS, viz., $9\times
1.5490$, needing a corresponding augmenting of the LHS, which can
come only from the IR terms from (2.9), symbolically denoted by
$[I'+ II']$ in (2.14), that include the (as yet free) parameter
$\lambda =2\beta \mu^2$.[Of course these IR terms do $not$
contribute to (2.12)].

\subsection{Dynamical Mass And Mass Renormalization}

The third point $p^2= - m_0^2$ which defines the dynamical mass,
corresponds to the ``pole' of the  propagator $S_F(p)$, so that
\begin{equation}\label{2.16}
 \Sigma^2 (i m_0) = + m_0^2 > \Sigma^2(0)
 \end{equation}
It may be recalled that a distinction between the dynamical and
constituent masses already exists in the literature. Thus in the
notation of Domb [29], (pp 322-324), $p$ and $p_0$ correspond to
$m_0$ and $\Sigma_0$ respectively].
\par
 Substituting from (2.11) gives a cubic equation in $m_0^2$:
\begin{equation}\label{2.17}
 \Sigma_0^2 = m_0^2 (1 - \beta m_0^2)^2
 \end{equation}
 which implies that $\Sigma_0 < m_0$.
Using the dimensionless variables $x_0 =\beta \Sigma_0^2 $ and
$y_0= \beta m_0^2$, this reduces to the cubic $y_0 (1-y_0)^2 =x_0$
which has at least one real solution for $y_0$ in terms of $x_0$:
\begin{equation}\label{2.18}
\beta m_0^2  \equiv y_0 = 2/3 + \sum_{\pm}[x_0 /2 -1/27 \pm
  \sqrt{x_0^2/4 - x_0/27}]^{1/3}
\end{equation}
whose nature can be seen as follows. For small $x_0$, $y_0$ is
also small (seen directly from the cubic form), but as $x_0$
increases, $y_0$ increases more rapidly, until $x_0$ reaches a
critical value $x_c = 4/27$ (seen from eq.(2.18)). Beyond this
point $y_0$ increases more slowly with $x_0$. The corresponding
critical value of $\beta$ is
\begin{equation}\label{2.19}
\beta_c = 4/(27 \Sigma_0^2 \approx 1.646 GeV^{-2}, for \Sigma_0 =
300 MeV
\end{equation}
We shall keep $\Sigma_0$ fixed at $300 MeV$, but vary $\beta$ in
the typical hadronic range $1.0 < \beta < 2.0$ for applications to
key QCD parameters like $<q{\bar q}>$ and $f_\pi^2$.  Now the
propagator may be written as
\begin{equation}\label{2.20}
 S_{FR}(p) = Z_m \frac{\Sigma(p)-i\gamma.p}{\Sigma^2(p) + p^2}
\end{equation}
making use of eq.(2.17), and formally introducing  a "mass
renormalization" factor $Z_m$ to be determined. However, using the
condition (2.16) in the numerator and denominator of (2.19) shows
immediately that near the pole the RHS already has the correct
structure $(m_0-i\gamma.p)/[m_0^2+p^2]$, which suggests that
$Z_m=1$ ! On the other hand an alternative way to extract the
factor $(p^2 + m_0^2)$ from (2.19), suggests a non-zero value of
$Z_m$. This is seen from rewriting the RHS of (2.20) as
$$
Z_m (\Sigma(p)-i\gamma.p)/[\Sigma^2(p)-\Sigma^2(im)+ m_0^2+p^2]
$$
and taking the limit $p^2 \rightarrow -m_0^2$ after extracting the
factor $(m_0^2 + p^2)$ from the denominator.  $Z_m$ is now
determined by the condition that at the $pole$ this quantity
reduce exactly to $1/(m_0+i\gamma.p)$. This gives
\begin{equation}\label{2.21}
Z_m = (1-3m_0^2 \beta) / (1- m_0^2 \beta)
\end{equation}
In view of this ambiguity in the working definition of $Z_m$, it
is not clear if this (finite) $Z_m$ is significant beyond unity.
However within this $subtractive$ renormalization approach to the
DSE, the divergences are already toned down to the logarithmic
level, so that renormalization is probably  less significant than
for the usual (unsubtracted) DSE form. For the rest therefore we
shall set $Z_m = 1$ in what follows.

\subsection{Solution  of  Eq.(2.9), Including IR Terms}

 Taking account of the  IR terms in (2.9) , the full equation (2.14) reads:
\begin{eqnarray}\label{2.22}
 0 &=& A x_0^2 - x_0 (B\lambda +1) + C \lambda  \\  \nonumber
 A &=& - \ln {4\pi} +\gamma +3 - 2\ln (\zeta / \Sigma_0)+ 9\ln(\zeta/\Lambda_Q)
   \nonumber  \\
 B &=& 7 - 6 \ln(\Sigma_0 / m_g); \quad C = 2- 2 \ln(\Sigma_0 / m_g)
\end{eqnarray}
where  $ x_0 = \beta \Sigma_0^2$, and $\lambda = 2\mu^2 \beta$. A
practical way is to solve (2.22)  for $\lambda$ with  $\Sigma_0 =
300 MeV $ and $m_g= 150 MeV$. This gives $\lambda$ for typical
values of the `range parameter $\beta$. Note that the connection
between $x_0$ and $y_0$ is already determined by (2.17-18). Now
with  $\Sigma=300 MeV$, a typical value  $\beta = 1 GeV^{-2}$,
($x_0 = 0.09$ and $y_0 = 0.115$), yields $\lambda =- 0.0640$. The
latter is an index of the strength of a (small)IR term needed to
provide a self- consistent solution of the DSE in the low momentum
regime to match its solution for `large' momenta. We shall come
back to these quantities in Sect.4 for the T-dependent DSE.
\par
At  this stage it may be asked as to what happened  to the  third
check-point $p^2= - m_0^2$ for the DSE, analogously to the points
$p^2 = \infty$  and $p^2 = 0 $ considered in the foregoing. As a
matter of fact, this condition has already been subsumed in the
determination of the relation between the constituent and
dynamical masses  in eqs.(2.17-2.18) within the specific structure
(2.11), so no new results can be expected from the DSE for $p^2 =
- m_0^2$. The check-point $p^2=-m_0^2$ will  however come into
play again in Sect.4, but in a T-dependent form of the DSE. But
before implementing the T-dependent DSE programme, it is first
necessary to carry out $two$ vital tests of this $T=0$ formalism,
viz., its performance on the two crucial quantities $<q{\bar q}>$
and $f_\pi^2$  [1,5, 6] which we consider next.

\section{Tests of Mass Function: $<q{\bar q}>$ And $f_\pi^2$}

\setcounter{equation}{0}
\renewcommand{\theequation}{3.\arabic{equation}}

The quark condensate and the  pion decay constant are regarded as
fairly sensitive tests of the mass function $\Sigma(p)$ determined
as a solution of the DSE, expressed in the differential form
(2.9). To that end we first collect their formal definitions as
follows. The condensate after tracing out the Dirac matrices is
\begin{equation}\label{3.1}
< q{\bar q}>_0 = \frac{ 4 N_c }{(2\pi)^4} \int d^4 p
\frac{\Sigma(p)} { \Sigma^2(p)  + p^2}
\end{equation}
which simplifies on making use  of  eq.(2.19-2.20) to
\begin{equation}\label{3.2}
<q{\bar q}>_0 = \frac{ 4 N_c }{(2\pi)^4} \int d^4 p \frac{\Sigma_0
(1+x)} {(p^2 + m_0^2) [(1+x)^2 -2y_0 (1 - y_0)]},
\end{equation}
where
\begin{equation}\label{3.3}
x = p^2 \beta; \quad y_0 = m_0^2 \beta; \quad x_0 =
\Sigma_0^2\beta
\end{equation}
The corresponding quantity $f_\pi^2$ may be defined in the chiral
limit by
$$ 2f_\pi^2  P_\mu = \frac{ N_c}{(2\pi)^4} \int d^4 p
Tr[(\Sigma(p_1) +\Sigma(p_2))\gamma_5 S_{FR}(p_1) i\gamma_\mu
\gamma_5  S_{FR}(-p_2)] $$ \\
where $S_{FR}$ is given by (2.19-20), and $ P= p_1 + p_2$, and the
pion-quark vertex function has been taken as [10] $[\Sigma(p_1) +
\Sigma(p_2)]/ 2f_\pi]$. Fortunately the complete expression may be
taken over from ref.[10] (also given in [1]), viz.,
\begin{equation}\label{3.4}
 f_\pi^2 = \frac{4 N_c}{(2\pi)^4} \int d^4 p_E \frac{[1-(p^2/4)\partial_{p^2}]\Sigma^2(p)}{(\Sigma^2(p) + p^2 )^2}
\end{equation}
in the Euclidean limit. The derivations of (3.2) and (3.4) are
shown in Appendices A and B respectively. The final result for the
condensate is summarized in (A.4-A.6) where the standard Table of
integrals [30] has often been employed. Similarly, for the pion
decay constant, the final result is given  by (B.5).

\subsection{Results for Condensate And Pion Decay}

The key parameters of this theory are $\Sigma_0$, the constituent
mass, and $\beta$, the parameter for the non- perturbative
hadronic scale. A third quantity, the dynamical mass $m_0$ is
determined by these via eq.(2.17), which can be expressed in terms
of the dimensionless parameters $x_0$) and $y_0$. Since the object
of this investigation is not to provide a detailed
phenomenological fit to these quantities, rather to see if this
new differential form of the DSE is consistent with the
conventional range of values of the $constituent$ mass, we shall
refrain from any fine-tuning and offer some typical values within
this alternative DSE framework, which is constrained by the fairly
rigid connection between $\Sigma_0$ and $m_0$ brought about by the
cubic equation (2.17). Thus, with a fixed $\Sigma_0$ at $300 MeV$,
Table 1 depicts some typical values of $\beta$, $x_0$ and $y_0$

\begin{center}
\Large \bf
Table I: Variations of $x_0$, $y_0$ With $\beta$ \\
\vspace{0.3in} \large
\begin{tabular}{|r|l|c|c|}
\hline
$\beta$&$x_0$&$y_0$ \\
\hline
1.00&0.090&0.115 \\
1.646&4/27&4/3 \\
2.00&0.135&1.365 \\
\hline
\end{tabular}
\end{center}

For these 3 sets we get under the MS scheme [28]
\begin{equation}\label{3.6}
<q{\bar q}>_0 = (0.1545; 0.0932; 0.114) \Sigma_0/ \beta
\end{equation}
where we have depicted the sensitivity of this quantity to the
main parameters $\beta$ for $\Sigma_0$ fixed at $300 MeV$. For the
values listed in Table I, the numbers work out at
$$(359 MeV)^3 ; (279 MeV)^3; (284 MeV)^3  $$
respectively, suggesting that $\beta$ should lie fairly close to
its `critical' value $\beta_c = 1.646$, without further tuning.
Similarly, the pionic constant works out for the 3 values of $x_0$
given above, as
\begin{equation}\label{3.5}
f_\pi^2 = (92.0 MeV)^2; (93.1 MeV)^2; (94.3 MeV)^2
\end{equation}
respectively, with $\Sigma_0 = 300 MeV$. This quantity is not
sensitive to $\beta$ but varies as the square of $\Sigma_0$.
 These values give a rough test of this formalism without
 vastly extending the numerical framework. Note that the
 the IR parameter $\lambda$ at $-0.064$ has been rather passive
 in these determinations, but its temperature dependence is
 going to play a more active role in the $T$-dependent  DSE,
for  a self-consistent determination of the critical temperature
$T_c$ to be considered in Sect.4 to follow.

\section{ T-DSE In Real Time Formalism}

\setcounter{equation}{0}
\renewcommand{\theequation}{4.\arabic{equation}}

As noted in Sect.(1.1), since our DSE formulation departs from the
more conventional thermodynamic formulations [20, 15] based on the
free energy [15] or effective potential [20], we are not yet in a
position to offer a complete set of critical indices near $T_c$,
except the one for the $T$-dependence of the order parameter
$m_t$. Keeping this in mind, to formulate the T-dependent DSE, we
have two broad options: real [22] vs imaginary [23] time
formalisms. The $T=0$ structure of the DSE suggests that it is
natural and convenient to employ the real time formalism and
follow the prescription of Dolen-Jackiw [22] for adding to the
quark and gluon propagators (which can be easily read off from the
main DSE, eq.(2.9)), the $T$ - dependent imaginary parts of the
Bose / Fermi types, leading to the modified propagators
respectively as follows
\begin{equation}\label{4.1}
D_{FT}(k) = \frac{- i}{k^2 + m_g^2} + \frac{2\pi}{\exp{\omega/T}-
1}\times \delta(k^2 + m_g^2); \quad \omega \equiv \sqrt{m_g^2 +
{\vec k}^2}
\end{equation}
\begin{equation}\label{4.2}
S_{FT}(p)= \frac{-i}{\Sigma(p)+ i\gamma.p}-
\frac{2\pi(\Sigma(p)-i\gamma.p)} {\exp{E_p/T}+1}\delta(\Sigma^2(p)
+ p^2)
\end{equation}
where the quark energy $E_p$ is the  fermionic analog of the gluon
energy $\omega$, eq.(4.1). Taking the gluon case first, there are
now two kinds of operations on (2.9). Namely, since the $p^2$
values are being considered on the mass shell, we shall now write
$p^2 = -m_t^2$ (instead of $-m_0^2$)to emphasize the T-dependence
of this quantity. Similarly (see Sect.1) we shall consider the
gluon mass $m_g$ and the constituent mass $\Sigma_0$ to ``run''
with $T$, and designate them as $\Sigma_t$ and $m_{gt}$
respectively. Considering the Bosonic and Fermionic Boltzmann
factors (4.1-4.2)in this order, we shall have extra contributions
to the four pieces on the RHS of (2.9), but giving rise to 3D
integrals only. We now collect these values separately, first
upgrading  the $T=0$ results of Sect 2 to $T \neq 0$.

\subsection{ T-Dependent $I; II$ and $I'; II'$}

To simplify the 4 pieces of the DSE, eq.(2.9), on the T-dependent
mass shell, the following results are useful.
\begin{equation}\label{4.3}
\Sigma(p)= m_t; \quad 2\Sigma'(p)-\Sigma''(p) = + \frac{\beta
m_t^4}{\Sigma_t};
\end{equation}
\begin{equation}\label{4.4}
 \Sigma(p) - \Sigma'(p)/2 = m_t - \frac{\beta m_t^4}{2 \Sigma_t}
\end{equation}
 Collecting these results on the (now T-dependent) 4 pieces on
the RHS of (2.9) we have
\begin{eqnarray}\label{4.5}
I+I'   &=& \frac{+\beta m_t^4}{\Sigma_t}[2.9538 +
\ln{(2\zeta^2/m_t^2)}
+\frac{2\lambda_t}{m_t^2\beta}[\ln{(m_t^2/2m_{gt}^2)}-1]] \\
\nonumber II+II' &=& (m_t-\frac{\beta
m_t^4}{2\Sigma_t}(2-\frac{8\beta m_t^3}{\Sigma_t}) [1 +
\frac{2\lambda_t}{m_t^2\beta}[\ln{(m_t^2/2m_{gt}^2)}-2]]
\end{eqnarray}

To these pieces must be added the $T$- parts of the gluon
propagators (Bosonic) accruing from (4.1), and the $T$ parts of
the quark propagators (Fermionic) from (4.2). These are basically
3D integrals because of the $\delta$-functions. To evaluate them
the following quantities come into play
\begin{equation}\label{4.6}
D(p-k) = \Sigma^2(p) + (p-k)^2 =- m_{gt}^2 + 2 m_t \omega
\end{equation}
\begin{equation}\label{4.7}
 4\Sigma(p)\Sigma'(p)+ 2p^2 -2 p.k = + \frac{4\beta m_t^5}{\Sigma_t}
-2 m_t^2 + 2 m_t \omega
\end{equation}
Here we have taken the rest frame of $p_\mu$, viz., ${\vec p}=0$.
The Bosonic $T$- parts normalized  to the pieces in (4.5) are
\begin{equation}\label{4.8}
BOSE_T = 4\int d\omega \sqrt{\omega^2
-m_{gt}^2}\frac{1}{\exp{\omega/T}-1} [ I_{BT} + II_{BT}]
\end{equation}
where the lower limit of $\omega$ integration is $m_{gt}$ and the
two integrands are
\begin{equation}\label{4.9}
I_{BT} = - \frac{\beta m_t^4/ \Sigma_t}{2\omega m_t - m_{gt}^2};
\end{equation}
\begin{equation}\label{4.10}
II_{BT}= \frac{ m_t - \frac{\beta m_t^4}{2 \Sigma_t}} {(2\omega
m_t - m_{gt}^2)^2} [2 m_t^2 -2 m_t\omega - 4m_t^5\beta/\Sigma_t]
\end{equation}
Similarly for the Fermionic parts, denoted by $I_(FT)$ and
$II_(FT)$ respectively.  The complete T-dependent DSE is then
obtained by modifying (2.9) a la (4.5) and adding the pieces
(4.9-4.10), and the corresponding Fermionic parts, after
integrations. Before carrying out these integrations we notice
some general features of these quantities in the neighbourhood of
the critical temperature $T_c$. Namely, \\
i) the powers of $m_t$ are spaced by $three$ units; \\
ii) $m_t$ and $m_{gt}$ are always involved in identical ratios.\\
One may infer from this  that  the critical index $\gamma$ for
both is the same at $3\gamma =1$, consistent with universality
[29] for such quantities. Thus in the neighbourhood of $T_c$ one
may take
\begin{equation}\label{4.11}
[m_t ; m_{gt}] \approx [\Sigma_0; m_g ] \tau^\gamma; \quad \tau
=1- T/T_c; \quad \gamma = 1/3
\end{equation}
The the $T$-dependence of $m_0$ may be handled via (2.17). Note
that in the neighbourhood of $T_c$, $\Sigma_t \approx m_t$, a
result which is consistent with eq. (7.278) of ref [29]. Retaining
only the lowest powers of the small quantities $m_t, m_{gt}$, most
of the terms in the $T-DSE$ will drop out, and the integrals over
(4.9-10) will lead to the net bosonic contribution
\begin{equation}\label{4.12}
BOSE_T /(4T) = \frac{m_t}{2m_{gt}(1-m_{gt}^2/2m_t^2)}- [\ln(2T/
m_{gt})]/2
\end{equation}
To this T-dependent (gluon propagator) contribution, must be added
the corresponding quark propagator contribution, eq.(4.2), near
$T=T_c$, by following a similar procedure to above. For brevity,
we indicate only the extra features, before writing the final
result. The Fermionic $T$- part of the quark propagator in(4.2)
now becomes
\begin{equation}\label{4.13}
(-2i\pi \frac{\delta((p-k)^2+ m_t^2)}{(\exp(E({\vec{p-k}}/T)+1))}
\end{equation}
And analogous to (4.6),
\begin{equation}\label{4.14}
k^2 + m_g^2 \approx 2 m_t E_k -2 m_t^2 + m_{gt}^2; \quad E_k =
\sqrt{({\vec k}^2 + m_t^2)}
\end{equation}
Next, taking account of eqs.(4.3-4), and proceeding as in the
gluon case, we can evaluate the quark counterpart of eq.(4.8) in
the neighbourhood of $T=T_c$ in the form
\begin{equation}\label{4.15}
 FERMI_T \approx [-T \beta m_t^2 \tan^{-1}[\frac{T}{T+m_t}]
 +\lambda_t(-\gamma + \ln(T/m_t))/4] /(4\pi^2)
 \end{equation}
It is easily checked that this quark contribution is at least of
$O(\sqrt{\beta}m_t)$ compared with the gluon's, so that it is
justified to neglect it, at least near the Critical point.  The
master equation $T-DSE$, keeping only the lowest order terms, now
simplifies to
\begin{equation}\label{4.16}
 \frac{4\lambda_t}{\beta m_t} L_1 + BOSE_T = 0;
\quad L_1 = \ln{(m_0^2/2 m_g^2)}-2
\end{equation}
This equation suggests a simple structure for $\lambda_t$, perhaps
one of the few that are consistent with its solution, viz.,
\begin{equation}\label{4.17}
\lambda_t = \lambda_0 (m_{gt}/m_g)^\gamma [-\ln\tau^\gamma +1] ;
\quad \tau = 1- T/ T_c
\end{equation}
where $\lambda_0$ may be identified with the value found in
Sect.3, viz., $\lambda =-0.064 \pm 0.003$, and the term unity in
square brackets signifies its normalization at $T=0$.  Eq.(4.16)
after substitution from (4.12), now reduces to $two$ equations,
involving the coefficients of
$$\tau^\gamma ;  \quad \tau^\gamma \ln {\tau^\gamma} $$
 respectively, but we skip these equations for brevity. The result,
 after elimination of the quantity $L_1$
 of (4.16) from them, and dividing out by $T$, is
\begin{equation}\label{4.18}
- 1/2 + (1/2\nu)[1-\nu^2/2] = 1/2 \ln[2T_c/m_g]; \quad
\nu=m_g/\Sigma_0
\end{equation}
using (4.11) near the Critical Point. Substituting from Sect.3.3,
viz., $\nu \approx 1/2$ gives the surprisingly simple result
\begin{equation}\label{4.19}
\nu \approx 1/2; \quad 2T_c \approx m_g \exp{7/8}
\end{equation}
leading to a reasonable value for the critical temperature, viz.,
\begin{equation}\label{4.20}
T_c \approx 180 \pm 20 MeV; \quad (m_g = 150 MeV).
\end{equation}

\subsection{Condensate and pionic constant near $T=T_c$}

For completeness we offer some brief comments on the predictions
of this simple formalism on the corresponding $T$-dependent
quantities $<q{\bar q}>$ and $f_\pi^2$ near the critical point
$T=T_c$, analogously to  the results of ref [1,5]. This is
possible in view of the $analytical$ expressions for these
quantities as given in Appendices A and B in terms of $y_0, a$ and
$x_0$ respectively. In $T$-dependent form, $y_0 \sim m_t^2, a \sim
m_t$, and $\Sigma_0 \sim m_t$, which in turn are expressible in
terms of the basic `order' parameters $m_t$ and $m_{gt}$,
eq.(4.11). Substitution in eqs.(A.6) and (B.5) shows that $ <q
{\bar q} > $ and $f^{2}_{\pi}$ tend to zero neat the Critical
Point like $m_t$ and $m_t^2 \ln{1/m_t^2}$ respectively, in general
accord with standard expectations.
\par
 For comparison with other approaches, the
chiral perturbation theory [6] for $f_\pi^(T)$ predicts [5]
$$  f_\pi (T)  = {\tilde f}_\pi [1 - \frac{N_f}{ 2 {\tilde f}_\pi^2 (2\pi)^3} \int d^3 p [\exp{(E/T)} -1]^{-1} /E] $$
which however does not indicate how this quantity behaves near
$T_c$.  Another form [1] which is more in line with our
parametrization of $\Sigma(p)$, suggests that $\Sigma_t$ should
vary as $<q{\bar q}>_T$, in agreement with our result for the
condensate.

\section{ Summary and Conclusion}

\setcounter{equation}{0}
\renewcommand{\theequation}{5.\arabic{equation}}

In retrospect, we have proposed a new (differential) form of the
Dyson-Schwinger Equation (DSE) for the mass function $\Sigma(p)$,
based on an (infinitesimal) $subtractive$ form of Renormalization
in QCD. Such `subtraction' in turn amounts to employing a
differential operator  of the form $p_\mu\partial_\mu $ applied on
both sides of the DSE, whose effect on the RHS is to $lower$ the
degree of divergence w.r.t. the integration variable $k_\mu$  by
$TWO$ units. It is in  the background of this (differential form
of) subtractive renormalization, that  it becomes possible to
approximate the quantity $\Sigma(p-k)$ inside the integral by
$\Sigma(p)$ since the $k$-dependence of this already decreasing
quantity is no longer sensitive  to the principal term $(p-k)^2$
in the quark propagator. [Without this background of an improved
$k$-convergence, however, this  approximation would not have been
justified ]. This crucial step which has facilitated the
integration over $d^4 k$ without further ado, has thus helped
convert the DSE into a $second-order$ differential equation, the
extra order (beyond the rainbow approximation [9]) arising from
the term responsible for satisfying  the WT identity a la
Pagels-Stokar [10], so as to preserve gauge invariance. To
reinforce this effect,  we have employed the Landau gauge which
makes the DSE virtually depend only on the mass function
$\Sigma(p)$ by effectively eliminating the $A$-function [27]. (The
`ghost terms' do not appear in this effective description).
\par
To solve the resulting differential form of the DSE, we have taken
recourse to $three$ crucial check-points: $p^2 = \infty$, $p^2 =
0$, and $p^2 = -m_0^2$,  using a pole ansatz, (2.16), (c.f. [1])
which is consistent with the form $p^{-2}$ in the large $p^2$
regime, in agreement with dynamical breaking of chiral symmetry
for massless quarks [10], provided the argument of $\alpha_s$ is
held fixed at some chosen value (here the UV parameter $\zeta$).
This has given a rather small value on the UV parameter $\zeta$
that appears as an argument of $\alpha_s$, which effectively
restricts the range of applicability of this formalism to moderate
values of $p^2$ (perhaps adequate for the cosmological application
envisaged in this paper). For the low $p^2$ regime, we have
introduced two kinds of masses: the $constituent$ mass $\Sigma(0)$
which is generally believed to be of $\sim 300 MeV$, and the
$dynamical$ mass $m_0$ which satisfies the equation $\Sigma(i m_0)
= m_0$ corresponding to the pole position of the quark propagator
$p^2 = - m_0^2$ (see also [29]). Now for the simple form (2.16)
the connection between the two `masses' is given by (2.18) (as the
solution of a cubic equation) which corresponds to $\Sigma(0) <
m_0$. The parameter $\beta$ in eq.(2.16), for a given $\Sigma(0)$,
has been taken as a typical hadronic scale befitting the low
energy regime of the DSE. The small IR parameter $2\mu^2 (\equiv
\lambda / \beta)$ which has played a passive role  in the $T=0$
description of the DSE, turns out to be rather crucial for $ T >
0$, for which the ansatz (4.17) is necessary for a self-consistent
solution of the $T-DSE$ (see further below). We have also
considered a non-zero value of the gluon mass for which several
arguments have been advanced in the literature [13-15].
\par
We have also carried out two important applications of $\Sigma(p)$
obtained from this new formulation of the DSE, viz., the quark
condensate and the pion decay constant. more by way of some basic
calibration of the formalism than as a means of detailed
phenomenological fits to hadronic data.  Thus a fit to within $<10
\%$ has helped fix  the parameters involved. After this check, we
have attempted in Sect.4  a $T$-dependent formulation of the DSE
to see the extent to which it can simulate the critical
temperature and at least one of the critical indices. To that end,
we have taken the $p^2 = -m_t^2$ limit of the T-DSE near the
Critical Point $T_c$ where it is small. In this respect, the
demands of consistency  have necessitated a $T$-dependence of the
IR confining parameter $\lambda$, for which an ansatz of the form
(4.17), calibrated to its value at $T=0$, is indicated. Two clear
results have emerged from the analysis, viz., i) a bunching of the
powers of $m_t$ in units of $three$ suggest a critical index
$\gamma = 1/3$ according to conventional analysis [29]; ii) and
the `matching' of the coefficients of like powers of the reduced
temperature $\tau$ have led to a very simple solution of the form
(4.18), leading  to the reasonable  $T_c$ at $180 \pm 20 MeV$.
\par
For a comparison of this result with those of contemporary
approaches [20,15], our approach  differs from these in one
important respect: the role of an external $H$-field is sought to
be partially simulated by the IR parameter $\lambda$ which is
necessarily $T$-dependent, instead of by  small but non-zero $u-d$
masses [20,15]. Further, in view of our explicit analytical
expressions for $<q{\bar q}>$ and $f_\pi^2$, we have also obtained
analytic structures for their $T$-dependence, and found indeed
that they both vanish at the critical point, without a detailed
numerical analysis [6,15, 20]. However this approach has its weak
points, especially the ad hoc nature of eq.(4.17) for the
$T$-dependence of the IR term. A second one is lack of a more
plausible understanding of the extent to which the IR term can
substitute for the current masses [20, 15] to simulate the
$H$-field effect. Attempts at throwing more light on these issues,
as well as extending the T-DSE formalism to facilitate the
evaluation of other critical indices [20, 29], are envisaged. And
in view of its central role, several other applications of the
"mass function", such as $\pi \rightarrow 2\gamma$, and e.m. pion
form factor at finite temperature [31], are under way.

\section{Acknowledgement}
This work was supported in part by the Taiwan CosPA Project of
Ministry of Education (MoE 89-N-FA01-1-4-3) and in part by
National Science Council (NSC90-2112-M-002-028). One of us (ANM)
is indebted to Olivier Pene for some useful comments.

\section{Appendix A: Evaluation of $<q{\bar q}>_0$ }

\setcounter{equation}{0}
\renewcommand{\theequation}{A.\arabic{equation}}

Using the notations $x=\beta p^2$ and $y_0 = \beta m_0^2$, and
anticipating a UV divergence  which requires a DR treatment [27],
we write  $4 \rightarrow n$ in eq.(3.2)  which reduces after the
angular integration [27, 11] to
\begin{equation}\label{A.1}
<q{\bar q}>_0 =  \frac{4 N_c \pi^{n/2} \Sigma_0 \zeta^\epsilon}
{(2\pi)^n  \Gamma(n/2) \beta^{n/2 - 1}} \int _0^\infty  d x
                        x^{n/2 - 1} F(x)
\end{equation}
where
\begin{equation}\label{A.2}
 F(x) = \frac{(1+x)}{(x + y_0)[(1+x)^2 -a^2]}; \quad a^2
=2y_0(1-y_0)
\end{equation}
Now break up $F(x)$ into partial fractions
 $$ F(x) = \frac{1}{[(1-y_0)^2- a^2]}[\frac{1-y_0}{x+y_0}-
 \frac{1-y_0+a}{2(1+x-a)}-\frac{1-y_0-a}{2(1+x+a)}] $$
The integration of each term above is carried out according to
\begin{equation}\label{A.3}
\int_0^\infty x^{n/2-1} dx/(A+x) =
A^{1-\epsilon/2}\Gamma(n/2)\Gamma(1-n/2)
\end{equation}
The rest is a matter of collecting all the 3 terms after giving a
DR [27] treatment to each. The final result is
\begin{eqnarray}\label{A.4}
<q{\bar q}>_0 &=& \frac{\Sigma_0 N_c }{\beta 4\pi^2} [g(a)(\ln
4\pi-\gamma+1+\ln\zeta^2\beta)+ h(a)];    \nonumber \\
g(a)          &=& \frac{1-y_0-a^2/2}{(1-y_0)^2- a^2}; \\ \nonumber
h(a)          &=& \frac{y_0(1-y_0)\ln
y_0}{(1-y_0)^2-a^2}-(1/2)\sum_\pm \frac{(1\pm a)\ln (1\pm
a)}{1-y_0 \pm a}
\end{eqnarray}
This result is valid for small $y_0$ (i.e., $\beta \sim 1$), when
$a^2 > 0$. However for larger $\beta$, vide eqs(2.18-19) of text,
$y_0$ exceeds unity, and $a^2 < 0$. For such cases, put $a^2 = -
b^2$. In particular the partial fraction break-up for $\beta_c$,
corresponding to $y_0 = 4/3$, is rather simple:
$$ F(x) = \frac{b^2 + (1+x)/3}{(1+x)^2 + b^2} - \frac{1/3}{x+y_0}
$$
since $b^2 + (y_0-1)^2$ becomes unity. Now using the result [30]
\begin{equation}\label{A.5}
\int_0^\infty \frac{x^{n/2-1}dx}{1+b^2+2x+x^2} =
-\frac{(1+b^2)^{n/4-1}\sin{(n/2-1)t}}{\sin t \sin{n\pi/2}}
\end{equation}
where
$$ \cos t = +1/\sqrt{1+b^2}; \quad \sin t = + b/\sqrt{1+b^2} $$
and giving a DR treatment [27] as above the corresponding result
to (A.5) is
\begin{eqnarray}\label{A.6}
<q{\bar q}>_0 &=& \frac{\Sigma_0 N_c }{\beta 4\pi^2}[b^2(\ln
4\pi-\gamma +1 +\ln\zeta^2\beta)+ f(b)]; \nonumber  \\
f(b)          &=& (1/6 -b^2/2)\ln(1+b^2)- \ln y_0 /3
-(4b/3)\tan^{-1}{b}
\end{eqnarray}

For purposes of obtaining the temperaure dependence of the quark
condensate (to be discussed in Sect.4 of the text), we record the
results of these integrations in the limit of small $a$ and $y_0$,
for which eq.(A.4) is appropriate:
\begin{equation}\label{A.7}
g(a) \approx 1 + O (y_0); \quad h(a) \approx y_0\ln y_0 + a^2 \sim
a\ln a
\end{equation}
Substitution in (A.1) gives in this limit
\begin{equation}\label{A.6}
<q{\bar q}>_0 = \frac{\Sigma_0 N_c }{\beta 4\pi^2}[(\ln
4\pi-\gamma+1+\ln\zeta^2\beta)+ O(a\ln a)]
\end{equation}
which lends itself immediately to a finite $T$ treatment in the
neighbourhood of the Critical point (see text, Sect.4).

\section{Appendix B: Evaluation of $f_\pi^2$}

\setcounter{equation}{0}
\renewcommand{\theequation}{B.\arabic{equation}}

Since the integral (3.4) is convergent by itself, DR [27] is not
needed in this case. After the angular integrations (using the
dimensionless units $x,y_0$ as before ), and carrying out the
differentiations, (3.4) reduces to
\begin{equation} \label{B.1}
f_\pi^2     =  \frac{\Sigma_0^2 N_c}{4\pi^2}{cal I}
\end{equation}
where the integral is defined by
\begin{equation}\label{B.2}
{\cal I}  = \int_0^\infty dx x\frac{(1+x)(1+3x/2)} {[x_0
+x(1+x)^2]^2}
\end{equation}
Now transform the variable from $x$ to $u$, as

$$  u  = \frac{x}{1+x}; \quad 0\leq u \leq 1. $$
The result of this is to give an integral in $u$ as
\begin{equation}\label{B.3}
{\cal I} = \int_0^1\frac{du u(1-u)(1+u/2)}
 {[x_0 (1-u)^3 + u]^2}
\end{equation}
While this integral is i principle exactly doable, it is
instructive to obtain an approximate $analytical$ expression which
in practice is sufficiently accurate, so as to lend itself to a
generalization to $finite$ temperatures (see below). The trick
lies is the observation that most of the contributions arise from
the region of $small$ values of $u$. Then (B.3) simplifies to
$$
{\cal I}\approx  \int_0^1\frac{du u(1-u/2)}
 {[x_0 (1-3u)+ u]^2}
 $$
Now integration by parts gives the final result
\begin{equation}\label{B.4}
{\cal I} = \frac{1/2}{(1-3x_0)^2}\ln{(1-2x_0)/x_0} -
\frac{1/2}{(1-2x_0)(1-3x_0)}
\end{equation}
Unlike the case of $<q{\bar q}>$, this result is valid for all
allowed $x_0$.  For purposes of determining the temperature
dependence of $f_\pi^2$, to be discussed in Sect.4, we record as
in Appendix A, the corresponding results for small $x_0$  This
gives
$$ {\cal I} \approx  1/2\ln{1/x_0} -1/2  $$
Substitution in (B.1) leads to
\begin{equation}\label{B.5}
f_\pi^2 \approx \frac{\Sigma_0^2 N_c}{8\pi^2
(1-3x_0)^2}[\ln{1/x_0} -\frac{1-3x_0}{1-2x_0}]
\end{equation}
which lends itself immediately to a finite $T$ treatment in the
vicinity of he Critical Point $T_c$ (see text, Sect.4).

\end{document}